\documentclass[twocolumn,letter]{jpsj2}
\def\e{{\rm e}}
\def\vector#1{{\bf #1}}

\def\vp{{\vector p}}
\def\vq{{\vector q}}

\def\dps{\displaystyle}

\def\Tc{{T_{\rm c}}}

\def\hightc{{high-$T_{\rm c}$ }}

\def\LSCO{\mbox{${\rm La_{2-{\it x}}Sr_{\it x}CuO_4}$}}

\def\SrRuO{\mbox{${\rm Sr_2RuO_4}$}}

\def\hsp#1{\hspace{#1ex}}

\def\Tc{{T_{\rm c}}}

\def\lsim{\stackrel{{\textstyle<}}{\raisebox{-.75ex}{$\sim$}}}
\def\gsim{\stackrel{{\textstyle>}}{\raisebox{-.75ex}{$\sim$}}}

\def\ppara{{p_{\parallel}}}

\def\omegaD{{\omega_{\rm D}}}

\def\eq.#1{eq.~(\ref{#1})} 
\def\pardif#1#2{\frac{\partial #1}{\partial #2}}

\newcommand\Equation[3]{
\begin{equation}\label{#1}\tag{#2}
#3
\end{equation}
}
\title{
Isotope Effect in Superconductors with Coexisting Interactions of \\
Phonon and Nonphonon Mechanisms 
}

\author{Hiroshi {\sc Shimahara}}
\def\runauthor{Hiroshi {\sc Shimahara}}

\inst{
Department of Quantum Matter Science, ADSM, Hiroshima University, 
Higashi-Hiroshima 739-8530, Japan
}

\recdate{May ~~, 2003}
\abst{
We examine the isotope effect of superconductivity in systems with 
coexisting interactions of phonon and nonphonon mechanisms 
in addition to the direct Coulomb interaction. 
The interaction mediated by the spin fluctuations is discussed 
as an example of the nonphonon interaction. 
Extended formulas for the transition temperature $\Tc$ and 
the isotope-effect coefficient $\alpha$ are derived 
for cases (a) $\omega_{\rm np} < \omegaD$ 
and (b) $\omega_{\rm np} > \omegaD$, 
where $\omega_{\rm np}$ is an effective cutoff frequency of 
the nonphonon interaction that corresponds to 
the Debye frequency $\omegaD$ in the phonon interaction. 
In case (a), it is found that the nonphonon interaction does not change 
the condition for the inverse isotope effect, 
{\it i.e.}, $\mu^{*} > \lambda_{\rm ph}/2$, 
but it modifies the magnitude of $\alpha$ markedly. 
In particular, it is found that a giant isotope shift occurs 
when the phonon and nonphonon interactions cancel each other largely. 
For instance, strong critical spin fluctuations may give rise to 
the giant isotope effect. 
In case (b), it is found that 
the inverse isotope effect occurs only when 
the nonphonon interaction and the repulsive Coulomb interaction, 
in total effect, work as repulsive interactions 
against the superconductivity. 
We discuss the relevance of the present result to 
some organic superconductors, such as $\kappa$-${\rm (ET)_2Cu(NCS)_2}$ 
and \SrRuO \hsp{0.25} superconductors, 
in which inverse isotope effects have been observed, 
and briefly to \hightc cuprates, 
in which giant isotope effects have been observed. 
}

\kword{
phonon-mediated pairing interactions, 
spin fluctuations, exotic superconductors, 
organic superconductor, \SrRuO, 
giant isotope shift, 
inverse isotope effect, 
negative isotope-effect coefficient
}

\begin{document}
%% JPSJ %% 
\sloppy
%% JPSJ %% 
\maketitle

%% PR %% FOR TWO COLUMN ACTIVATE THE LINE BELOW
%% ]

%% PR %% \narrowtext

The mechanisms of the pairing interactions of nonphonon origins 
have been examined for anisotropic superconductivity. 
For example, spin fluctuation exchange interactions have been 
examined for the superfluid $^{3}$He~\cite{And73} 
and exotic superconductors, 
such as 
heavy-fermion superconductors~\cite{Sca86}, 
organic superconductors~\cite{Eme86,Bea86,Shi89}, 
\hightc cuprates~\cite{Miy88}, 
and \SrRuO \hsp{0.25} superconductors~\cite{Ric95}. 
In order to clarify the origin of the pairing interactions, 
the isotope effect has been observed in many superconductors. 
It is widely known that the isotope-effect coefficient $\alpha = 0.5$ 
observed in nontransition-metal superconductors, such as Hg and Sn, 
is evidence of phonon-mediated pairing interactions in those 
systems. 
Here, the isotope-effect coefficient $\alpha$ has been defined by 
\Equation{eq:alphadef}
{1}
{
     \alpha = - \pardif{\ln \Tc}{\ln M} , 
     }
where $M$ is the relevant atomic mass. 
The value $\alpha = 0.5$ is directly derived from 
the BCS formula 
${
     \Tc = 1.13 \omegaD \e^{-1/\lambda} 
     }$ 
with $\omegaD \propto M^{-1/2}$ and $\lambda \propto M^{0}$.

However, deviation of $\alpha$ from $0.5$ is not necessarily 
evidence of the presence of the pairing interaction 
of a nonphonon origin. 
For instance, in transition-metal superconductors, such as Ru and Os, 
the value of $\alpha$ largely deviates from 0.5 
due to the strong repulsive Coulomb interaction $V_{\rm C}$. 
In the presence of it, $\lambda$ is replaced with 
${\tilde \lambda} \equiv \lambda-\mu^{*}$, where 
\Equation{eq:effectiveCoulombparameter}
{2}{
     \mu^{*} = \frac{\mu}{1 + \mu \ln (W/\omegaD)} , 
     }
$\mu \equiv V_{\rm C} N(0)$ 
and $W$ denotes an effective cutoff energy of the Coulomb interactions 
of the order of the band width. 
Hence, we obtain 
\Equation{eq:alphawithmu}
{3}{
     \alpha = \frac{1}{2} \Bigl [
              1 - \bigl ( \frac{\mu^{*}}{\tilde \lambda} \bigr )^2 
              \Bigr ] . 
     }
In particular, the strong Coulomb repulsion 
such as $\lambda > \mu^{*} > \lambda/2$ gives rise to 
the inverse isotope effect ($\alpha < 0$). 
The anharmonicity of the lattice vibrations may also affect 
the isotope effect in compounds, 
such as PdH(D) ($\alpha \approx -0.25$)~\cite{Kle92}. 
The inverse isotope effect has been observed also 
in some organic superconductors~\cite{Sch96,Kin96,IEorg,Hei86,Osh88,Ito91,Tok91} 
and \SrRuO~\cite{Mao01}, but the mechanisms is unknown.

In most cases, a nonzero isotope shift ($\alpha \ne 0$) suggests 
the presence of the phonon contribution to the pairing interactions. 
One might expect that isotope shift can be attributed to 
the change in normal state properties, such as the density of states. 
However, usually, such changes should be of the order of 
$(\Delta M/2M) (\omega_{\rm D}/ \epsilon_{\rm F})$, 
which is usually negligible except for very narrow band systems, 
where $\Delta M$ is the shift of the atomic mass. 
Therefore, the isotope effect is mainly attributed to a change 
in the pairing interactions.

Here, it could not be excluded {\it a priori} that $\alpha \ne 0$ occurs 
because the phonon interaction $V_{\rm ph}(\vp,\vp')$ 
reduces the anisotropic pairing interaction, but it seems unlikely. 
The effective interaction for the anisotropic gap function 
proportional to $\gamma(\vp)$ is essentially proportional to 
the average of 
$V_{\rm ph}(\vp,\vp') \, \gamma^{*}(\vp) \gamma(\vp')$ on the Fermi surface, 
where $\gamma(\vp)$ is a function for expressing the momentum dependence. 
Since $V_{\rm ph}(\vp,\vp')$ tends to take larger negative values 
for smaller momentum transfers, {\it i.e.}, for $\vp \approx \vp'$, 
where $\gamma^{*}(\vp) \gamma(\vp') > 0$, 
the above average must be negative. 
For example, 
in order to obtain a repulsive effective interaction 
for $\gamma(\vp) \propto p_x$, 
$V_{\rm ph}(\vp,\vp')$ must have a negative peak 
at $|\vp - \vp'| \sim 2 k_{\rm F}$ rather than $|\vp - \vp'| \sim 0$, 
but it is unlikely. 
In fact, some explicit calculations have shown that 
the phonon interaction includes attractive components 
for anisotropic pairing~\cite{Fou77,Shi02a,Shi02b}. 
Therefore, the magnitude of the isotope shift of $\Tc$ reflects, 
more or less, the degree of the positive contribution 
to the Cooper pair formation from the phonons.

We should note that spin triplet pairing does not necessarily exclude 
the phonon mechanisms of the pairing interaction. 
The triplet pairing interaction is usually weaker than 
the singlet pairing interaction in the phonon mechanism, 
but it could become dominant, for example, 
in the presence of a strong short-range Coulomb 
interaction~\cite{Fou77,Shi02a,Shi02b}, 
which strongly suppresses the isotropic pairing more than 
the anisotropic pairing. 
In this context, the large deviation of $\alpha$ from 0.5 
is a natural consequence of the strong Coulomb repulsion. 
The layer structure also favors the anisotropic pairing interactions 
mediated by phonons~\cite{Shi02b}.

In this paper, we examine the isotope effect in systems in which 
interactions of phonon and nonphonon mechanisms coexist. 
In particular, we examine the condition for the inverse isotope effect.

We assume that the nonphonon interaction also has an effective cutoff 
frequency $\omega_{\rm np}$ that corresponds to 
the Debye frequency $\omegaD$ in the phonon interaction. 
This assumption is explained as follows 
for the interactions mediated by the spin fluctuations. 
The spin fluctuations have a characteristic frequency $\omega_{\rm sf}$, 
which becomes smaller as one approaches the magnetic instability point 
for the critical slowing down. 
The structure of the susceptibility $\chi(\vq,\omega)$ reflects 
this characteristic frequency. 
The susceptibility $\chi(\vq,\omega)$ has a sharp peak 
around a momentum $\vq_0$ 
with a width corresponding to $\omega_{\rm sf}$, 
in addition to a broad peak over the whole momentum space, 
which reflects the electron characteristic energy $\epsilon_{\rm F}$. 
Since the effective interaction mediated by the spin fluctuations 
is essentially proportional to $\chi(\vq,\omega)$, 
it consists of two parts with different 
characteristic energies $\omega_{\rm sf}$ and $\epsilon_{\rm F}$. 
In the gap equation, these characteristic energies play roles of 
the effective cutoff energies of the two parts of the effective interaction. 
The part of the effective interaction with $\epsilon_{\rm F}$ 
can be included in the repulsive Coulomb interaction, 
which also has a large effective cutoff energy of the same order. 
Therefore, we may regard that 
the interaction mediated by the spin fluctuations 
has an effective cutoff energy of the order of $\omega_{\rm sf}$, 
namely, $\omega_{\rm np} \sim \omega_{\rm sf}$. 
In fact, it has been shown in a theoretical calculation~\cite{Shi89} 
that the contribution to the pairing interaction mainly comes from 
this sharp peak of $\chi$. 
In theoretical phase diagrams~\cite{Shi88,Shi89}, 
the superconductivity by such a pairing interaction occurs 
only near the magnetic phase boundary. 
An explicit calculation, which illustrates the above argument, 
has been given in ref.~\citen{Shi89}.

We start with the interactions between electrons, 
\Equation{eq:interaction}
{4}{
     V(\vp,\vp') 
     = \sum_{\alpha} 
       \sum_{k=1}^{n} V_{k}^{\alpha} (\xi_{\vp},\xi_{\vp'}) 
         \gamma_{\alpha}(\ppara) \gamma_{\alpha}(\ppara') , 
     }
where 
$\ppara$ and $\xi_{\vp}$ 
denote the two-dimensional momentum coordinate on 
the Fermi surface and the electron energy measured from the Fermi surface 
with indices $k$ and $\alpha$ 
to express the kinds of interactions and the pairing symmetries. 
The symmetry functions $\gamma_{\alpha}(\ppara)$ are normalized 
with respect to the average on the Fermi surface~\cite{Shi03}. 
We consider a gap function proportional to $\gamma_{\alpha}(\ppara)$, 
retaining only terms with $\alpha$ in \eq.{eq:interaction}. 
Introducing cutoff energies $\omega_k$, we put 
\Equation{eq:interactionwithcutoff}
{5}{
     V_k^{\alpha}(\xi,\xi') 
     = V_{k0}^{\alpha} \, 
          \theta(\omega_k - |\xi|) \, \theta(\omega_k - |\xi'|) , 
     }
where $\omega_1 < \omega_2 < \omega_3 < \cdots < \omega_n$. 
We should note that $V_{k0}^{\alpha}$ for the repulsive Coulomb interaction 
can be nonzero even for anisotropic pairing, 
unless we assume only on-site repulsion. 
For \eq.{eq:interactionwithcutoff}, 
the gap function is written as 
\Equation{eq:gapfunctionexpand}
{6}{
     \Delta_{\alpha}(\vp) 
     = \sum_{k=1}^{n} \Delta_{k \alpha} \, 
          \theta(\omega_k - |\xi_{\vp}|) \, \gamma_{\alpha}(\ppara) . 
     }
The linearized gap equation ($\Tc$ equation) is written as 
\Equation{eq:gapeq}
{7}{
     \Delta_{k \alpha}
     = \lambda_{k} 
          \Bigl [  
               \sum_{k' = 1}^{k-1} \Delta_{k' \alpha} l_{k'} 
             + \sum_{k' = k}^{n}   \Delta_{k' \alpha} l_{k} 
          \Bigr ] , 
     }
where we have defined $\lambda_{k} = - V_{k0}^{\alpha} N(0)$ and 
$l_{k} = \ln ({2 \e^{\gamma} \omega_{k}}/{\pi \Tc})$ 
with the Euler constant $\gamma = 0.57721\cdots$. 
We write the transition temperature in the form 
\Equation{eq:Tclambdatilde}
{8}{
     \Tc = \frac{2 \e^{\gamma}}{\pi} \omega_{1} \e^{-1/{\tilde \lambda}_1} , 
     }
defining the effective coupling constant ${\tilde \lambda}_1$. 
We define 
${
     \alpha_{k0} = - {\partial \ln \omega_{k}}/{\partial \ln M} 
     }$ 
for convenience.

First, we consider the case with $n = 2$. 
The $\Tc$ equation is written as 
\Equation{eq:n2casegapeq}
{9}{
     \left ( \hsp{-1} 
       \begin{array}{cc}
         \lambda_{1}^{-1} - l_1 & - l_1 \\
         - l_1 & \lambda_{2}^{-1} - l_2 \\
       \end{array}
     \hsp{-1} \right ) 
     \left ( \hsp{-1} 
       \begin{array}{c}
         \Delta_{1 \alpha} \\
         \Delta_{2 \alpha} \\
       \end{array}
     \hsp{-1} \right ) 
     = 
     \left ( \hsp{-1} 
       \begin{array}{c}
         0 \\ 
         0 \\ 
       \end{array}
     \hsp{-1} \right ) . 
     }
Therefore, we obtain 
${\tilde \lambda}_1 = \lambda_{1} + \lambda_2^{*}$ 
and 
$
     \lambda_2^{*} 
     = \lambda_{2}/[1 - \lambda_{2} \ln(\omega_2/\omega_1)]
     $. 
The isotope-effect coefficient is obtained as
\Equation{eq:alphan2case}
{10}{
     \alpha = \alpha_{10} 
              \Bigl [
              1 - \bigl ( \frac{\lambda_2^{*}}
                               {\lambda_{1} + \lambda_2^{*}} 
                  \bigr )^2 
              \Bigr ] 
              + 
              \alpha_{20} 
              \bigl ( \frac{\lambda_2^{*}}
                           {\lambda_{1} + \lambda_2^{*}} 
              \bigr )^2 . 
     }
If we take $V_{10}^{\alpha}$ and $V_{20}^{\alpha}$ 
as a phonon-mediated pairing interaction and 
the repulsive Coulomb interaction, respectively, 
then $\omega_1 = \omegaD \propto M^{-1/2}$ and 
$\omega_2 = W \propto M^{0}$, 
and \eq.{eq:alphan2case} is reduced to the standard formula 
(\ref{eq:alphawithmu}).

Equation~(\ref{eq:alphan2case}) can be used for the systems in which 
two different interactions coexist and 
the repulsive Coulomb interaction is negligible. 
Here, we consider a system in which one of the two interactions 
is of a nonphonon origin. 
We examine two cases: 
(a) $\omega_{\rm np} < \omegaD$ and 
(b) $\omegaD < \omega_{\rm np}$, 
where $\omega_{\rm np} \propto M^{0} $ and 
$\omegaD \propto M^{-1/2}$. 
If the nonphonon interaction is one mediated by the spin fluctuations, 
$\omega_{\rm np} \sim \omega_{\rm sf}$, 
and case (a) occurs in proximity to the magnetic instability. 
We define 
$\lambda_{\rm ph} \equiv - V_{\rm ph}^{\alpha} N(0)$ and 
$\lambda_{\rm np} \equiv - V_{\rm np}^{\alpha} N(0)$, 
where $V_{\rm ph}^{\alpha}$ and $V_{\rm np}^{\alpha}$ denote 
the coupling constants of the phonon and nonphonon interactions, 
respectively.

For case (a), we put 
$\omega_1 = \omega_{\rm np}$, 
$\omega_2 = \omegaD$, 
$\lambda_{1} = \lambda_{\rm np}$, 
and $\lambda_2^{*} = \lambda_{\rm ph}^{*} $. 
Thus, we obtain 
\Equation{eq:alphan2case01}
{11}{
     \alpha = \frac{1}{2}
              \bigl ( \frac{\lambda_{{\rm ph}}^{*}}
                           {\lambda_{\rm np}
                            + \lambda_{{\rm ph}}^{*}}
                  \bigr )^2 , 
     }
which is positive definite. 
On the other hand, for case (b), 
we put 
$\omega_1 = \omegaD$, 
$\omega_2 = \omega_{\rm np}$, 
$\lambda_{1} = \lambda_{\rm ph}$, 
and $\lambda_2 = \lambda_{\rm np} $. 
Thus, we obtain 
\Equation{eq:alphan2case02}
{12}{
     \alpha = \frac{1}{2}
              \Bigl [
              1 - \bigl ( \frac{\lambda_{\rm np}^{*}}
                               {\lambda_{\rm ph} 
                                    + \lambda_{\rm np}^{*}} 
                  \bigr )^2 
              \Bigr ] , 
     }
where $\lambda_{\rm np}^{*} 
= \lambda_{\rm np}/[1 - \lambda_{\rm np} \ln(\omega_{\rm np}/\omegaD)]$. 
Therefore, $\alpha$ could become negative only 
in case (b) ($\omega_{\rm np} > \omegaD$) with 
$\lambda_{\rm ph} \lambda_{\rm np}^{*} < 0$.

\vspace{\baselineskip}

%%%%%%%%%%%%%%%%%%%%%%%%%%%%%%%%%%%%%%%%%%%%%%%%%%%%%%%%%%%%%%%%%%%%%%%
%%  Fig.1                                                            %%
%%%%%%%%%%%%%%%%%%%%%%%%%%%%%%%%%%%%%%%%%%%%%%%%%%%%%%%%%%%%%%%%%%%%%%%
%% JPSJ %% 
\begin{figure}
%% JPSJ %% When you do not use epsf, activate the next line. 
%% \figureheight{7cm}
%% 
%% PR %% \begin{figure}[htb]
\begin{tabular}{c}
%% FOR TWO COLUMN 
%% \leavevmode \epsfxsize=6cm  
%% 
%% FOR ONE COLUMN 
%% \leavevmode \epsfxsize=10cm  
%% JPSJ2 %% 
\includegraphics[width=6.0cm]{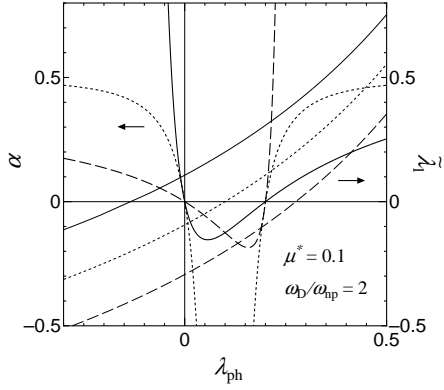}
%% PR %% \leavevmode \epsfxsize=7cm  
%% PR %% \epsfbox{fig1.eps}
\end{tabular}
\caption{
Isotope-effect coefficients as functions of $\lambda_{\rm ph}$ 
for case (a) $\omega_{\rm np} < \omegaD$. 
Solid, dashed, and dotted curves show the resuls for 
$\lambda_{\rm np} = 0.2$, $0$, and $-0.2$, respectively. 
Thick and thin curves show the values of $\alpha$ and ${\tilde \lambda}_1$, 
respectively. 
} 
\label{fig:case_a}
\end{figure}
%%%%%%%%%%%%%%%%%%%%%%%%%%%%%%%%%%%%%%%%%%%%%%%%%%%%%%%%%%%%%%%%%%%%%%%

Next, we examine the effect of the direct Coulomb interaction. 
The $\Tc$ equation with $n = 3$ 
can be solved in a similar manner to that in the above. 
It is obtained that 
${\tilde \lambda}_1 = \lambda_{1} + \lambda_2^{*}$, 
$\lambda_2^{*} = ({\lambda_{2} + \lambda_3^{*}})
               /[{ 1 - (\lambda_{2} + \lambda_3^{*}) \, l_{21} }]$, 
and $\lambda_3^{*} = {\lambda_{3}}
               /[{ 1 - \lambda_{3} l_{32} }]$, 
where 
$l_{kk'} \equiv \ln (\omega_k/\omega_k')$. 
The transition temperature is given by \eq.{eq:Tclambdatilde}. 
Hence, we obtain 
\Equation{eq:alphan3case}
{13}{
     \alpha = \alpha_{10} C_1 
            + \alpha_{20} C_2 
            + \alpha_{30} C_3 , 
     }
where 
\Equation{eq:C1etcdef}
{14}{
     \begin{array}{rcl}
     C_1 \hsp{-2} & = & \hsp{-2} \dps{ 
          1 - \bigl ( \frac{\lambda_2^{*}}
                           {\lambda_{1} + \lambda_2^{*}} \bigr )^2 , 
     } \\[10pt]
     C_2 \hsp{-2} & = & \hsp{-2} \dps{ 
          \bigl ( \frac{\lambda_2^{*}}{\lambda_1 + \lambda_2^{*}} \bigr )^2 
          \Bigl [ 
          1 - \bigl ( \frac{\lambda_3^{*}}
                          {\lambda_{2} + \lambda_3^{*}} \bigr )^2 
          \Bigr ] , 
     } \\[10pt]
     C_3 \hsp{-2} & = & \hsp{-2} \dps{ 
          \bigl ( 
          \frac{ \lambda_{2}^{*} }
               { \lambda_1 + \lambda_{2}^{*} } 
          \bigr )^2 
          \bigl ( 
          \frac{ \lambda_{3}^{*} }
               { \lambda_{2} + \lambda_3^{*} }
          \bigr )^2 . 
     } \\
     \end{array}
     }
We examine cases (a) and (b) mentioned above, 
adding the direct Coulomb interaction $V_{3}^{\alpha}$ 
with $\omega_3 = W$ and $\lambda_{3} = - \mu$.

In case (a), we obtain 
${\tilde \lambda}_1 = \lambda_{\rm np} + \lambda_{\rm ph}^{*}$ and 
\Equation{eq:alpha_for_case_a}
{15}{
     \alpha = \frac{1}{2} 
          \bigl ( \frac{\lambda_{\rm ph}^{*}}
                       {\lambda_{\rm np} + \lambda_{\rm ph}^{*}} \bigr )^2 
          \Bigl [ 1 - \bigl ( \frac{\mu^{*}}{\lambda_{\rm ph} - \mu^{*}} 
                      \bigr )^2 \Bigr ] , 
     }
where ${
     \lambda_{\rm ph}^{*} 
     = (\lambda_{\rm ph} - \mu^{*})
        /[ 1 - (\lambda_{\rm ph} - \mu^{*}) \, 
                   \ln (\omegaD/\omega_{\rm np}) ] 
     }$ 
with $\mu^{*}$ defined by \eq.{eq:effectiveCoulombparameter}. 
In this case, the sign of $\alpha$ is determined only by the ratio 
$\lambda_{\rm ph}/\mu^{*}$. 
If we assume $\mu^{*} > 0$, 
there is no region of $\lambda_{\rm ph}< 0$ where $\alpha < 0$ occurs. 
In Fig.~\ref{fig:case_a} and \eq.{eq:alpha_for_case_a}, 
it is also found that $\alpha$ rapidly varies 
with $\lambda_{\rm ph}$ and $\lambda_{\rm np}$. 
The prefactor 
$[{\lambda_{\rm ph}^{*}}/(\lambda_{\rm np} + \lambda_{\rm ph}^{*})]^2$ 
in \eq. {eq:alpha_for_case_a} 
gives rise to a large value of $\alpha$ 
when $\lambda_{\rm ph}^{*} \approx - \lambda_{\rm np}$.

\begin{figure}
%% JPSJ %% When you do not use epsf, activate the next line. 
%% \figureheight{7cm}
%% 
%% PR %% \begin{figure}[htb]
\begin{tabular}{c}
%% FOR TWO COLUMN 
%% \leavevmode \epsfxsize=6cm  
%% 
%% FOR ONE COLUMN 
%% \leavevmode \epsfxsize=10cm  
%% JPSJ2 %% 
\includegraphics[width=5.8cm]{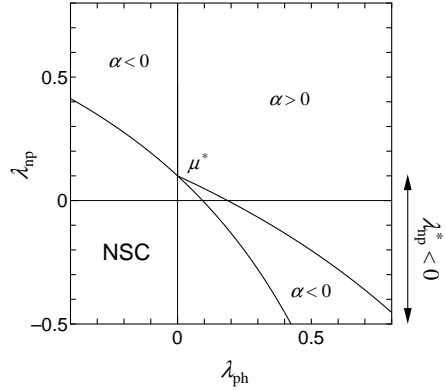}
%% PR %% \leavevmode \epsfxsize=7cm  
%% PR %% \epsfbox{fig2.eps}
\end{tabular}
\caption{
Phase diagram for case (b) $\omega_{\rm np} > \omegaD$, 
with $\omega_{\rm np}/\omegaD = 2$ and $\mu^{*} = 0.1$. 
NSC denotes the region where the superconductivity does not occur. 
} 
\label{fig:case_b}
\end{figure}

On the other hand, in case (b), we obtain 
${\tilde \lambda}_1 = \lambda_{\rm ph} + \lambda_{\rm np}^{*}$ and 
\Equation{eq:alpha_for_case_b}
{16}{
     \alpha = \frac{1}{2} 
          \Bigl [ 1 - 
               \bigl ( \frac{\lambda_{\rm np}^{*}}
                            {\lambda_{\rm ph} + \lambda_{\rm np}^{*}}
               \bigr )^2 \Bigr ] , 
     }
where 
${
     \lambda_{\rm np}^{*} 
     = (\lambda_{\rm np} - \mu^{*})
       /[ 1 - (\lambda_{\rm np} - \mu^{*}) \, 
                 \ln (\omega_{\rm np}/\omegaD) ] 
     }$ 
and in this case $\mu^{*} = \mu/[1 + \mu \ln (W/\omega_{\rm np})]$. 
It is found that $\lambda_{\rm ph} \lambda_{\rm np}^{*} < 0$ is 
a necessary condition for $\alpha < 0$. 
Since it is reasonable to assume $\lambda_{\rm ph} > 0$ 
even for anisotropic pairing as argued above, 
an inverse isotope effect suggests 
that the nonphonon interaction and 
the direct Coulomb interaction, in total effect, 
work as repulsive interactions 
against the superconductivity ($\lambda_{\rm np}^{*} < 0$) 
when $\omega_{\rm np} > \omegaD$. 
Figure~\ref{fig:case_b} shows the phase diagram obtained using 
eq.~(\ref{eq:alpha_for_case_b}).

\vspace{\baselineskip} 

%%%%%%%%%%%%%%%%%%%%%%%%%%%%%%%%%%%%%%%%%%%%%%%%%%%%%%%%%%%%%%%%%%%%%%%
%%  Fig.2                                                            %%
%%%%%%%%%%%%%%%%%%%%%%%%%%%%%%%%%%%%%%%%%%%%%%%%%%%%%%%%%%%%%%%%%%%%%%%
%% JPSJ %% 
%%%%%%%%%%%%%%%%%%%%%%%%%%%%%%%%%%%%%%%%%%%%%%%%%%%%%%%%%%%%%%%%%%%%%%%

Now, we describe the general solution of \eq.{eq:gapeq} 
with any positive integer $n$. 
Equation (\ref{eq:gapeq}) is rewritten as 
$\Delta_{k} = \lambda_{k} F_k$, where we have defined 
$F_k \equiv \sum_{k'=1}^{k-1} l_{k'} \Delta_{k'} + l_k {\bar \Delta}_k $ 
with ${\bar \Delta}_k \equiv \sum_{k'=k}^{n} \Delta_{k'}$. 
On the other hand, 
it is easily proved from eq.~(\ref{eq:gapeq}) with $k+1 \sim n$ 
that an equation of the form 
${\bar \Delta}_{k+1} = \lambda_{k+1}^{*} F_k$ holds for $k \geq 1$ 
with a recurrence formula 
\Equation{eq:lambdaast_rec}
{17}
{
     \lambda^{*}_{k} = \frac{\lambda_{k} + \lambda_{k+1}^{*}}
          {1 - (\lambda_{k} + \lambda_{k+1}^{*}) l_{k,k-1}} 
     }
for $k \geq 2$. 
Hence, it holds that 
${\bar \Delta}_1 = \Delta_1 + {\bar \Delta}_2 
= (\lambda_{1} + \lambda_{2}^{*}) l_1 {\bar \Delta}_1$. 
Therefore, $\Tc$ is given by \eq.{eq:Tclambdatilde} with 
${\tilde \lambda}_{1} = \lambda_{1} + \lambda_{2}^{*}$~\cite{Yam87}. 
Hence, from \eq.{eq:alphadef}, the general formula for $\alpha$ 
is obtained as 
\Equation{eq:alphageneral}
{18}
{
     \alpha = \sum_{k=1}^{n} C_{k} \alpha_{k0} 
     } 
with 
$C_{k} = \Lambda_{k} - \Lambda_{k+1}$ for $k \leq n-1$ 
and $C_{n} = \Lambda_{n}$, 
where $\Lambda_{k} = \Pi_{l=1}^{k-1} 
[\lambda_{l+1}^{*}/(\lambda_{l} + \lambda_{l+1}^{*})]^2$ for $k \geq 2$ 
and $\Lambda_{1} = 1$. 
%%  In the derivation, it is convenient to put 
%%  $\Delta {\tilde \alpha}_{k0} \equiv \Lambda_{k} \Delta {\alpha}_{k0}$ 
%%  and ${\Delta \alpha_{k0}} \equiv 
%%  \partial [(\lambda_{k}^{*})^{-1}]/\partial \ln M$. 
For example, if the interactions $V_{k}^{\alpha}$ with $k = 1,\cdots n-1$ 
are phonon interactions with $\alpha_{k0} = 1/2$, 
it is found that $\alpha \approx 1/2$ for large $n$, 
since $\Lambda_n \ll 1$.

Let us discuss the inverse isotope effect observed in clean samples of 
\SrRuO \hsp{0.25} with $\Tc \gsim 0.94 \Tc_0$~\cite{Mao01}, 
where $\Tc_0$ is the transition temperature in the clean limit. 
From the present result for case (b), 
which is probably the case for \SrRuO~\cite{Note03}, 
the inverse isotope effect suggests 
that the superconductivity in this compound is induced by 
the phonon-mediated pairing interaction~\cite{Note02}.

This result is consistent with the recent report 
that the antiferromagnetic fluctuation does not 
seem to play a constructive role in the triplet 
superconductivity in \SrRuO~\cite{Kik02}. 
The impurity concentration dependence of the isotope-effect 
coefficient~\cite{Mao01} can also be reproduced 
within a theory based on a phonon mechanism~\cite{Shi03}. 
Many novel mechanisms have been proposed for the superconductivity 
in this compound~\cite{Note01}. 
However, there is no conclusive evidence for the nonphonon mechanism 
at the present. 
The present theory does not consider the effect of anharmonicity 
of lattice vibrations. 
However, if such effect is essential for the inverse isotope effect, 
we also obtain the same conclusion that the phonon interaction 
is dominant.

The situation is rather different in the organic superconductors. 
The inverse isotope effect was also observed 
in the organic superconductors, such as 
$\kappa$-${\rm (ET)_2Cu(NCS)_2}$~\cite{Osh88,Kin96,IEorg}, 
$\beta$-${\rm (ET)_2I_3}$~\cite{Hei86,IEorg}, 
and $\kappa_{\rm L}$-${\rm (ET)_2Ag(CF_3)_4}$
(1-bromo-1,2-dichloroethane)~\cite{Sch96}, 
while a normal isotope effect was observed 
in $\kappa$-${\rm (ET)_2Cu[N(CN)_2]Br}$~\cite{Ito91,Tok91,IEorg}. 
In contrast to \SrRuO, case (a) ($\omega_{\rm sf} < \omegaD \lsim W$) 
may be realized in the organic superconductors 
even if $\xi_{\rm mag} = 2 a \sim 3 a$, 
because they have narrower band widths of $0.2{\rm eV} \sim 0.5{\rm eV}$. 
The coefficient $\alpha$ varies rapidly, as shown in Fig.~\ref{fig:case_a}, 
when $\lambda_{\rm ph}$ changes, 
whichever $\lambda_{\rm np} > 0$ or $< 0$. 
Thus, possibilities of the nonphonon pairing interactions 
are not excluded in the organic superconductors with $\alpha < 0$.

In \hightc cuprates \LSCO, 
it was observed by Crawford {\it et al.}~\cite{Cra90} 
that $\alpha$ increases near the antiferromagnetic boundary, 
where $\Tc$ decreases. 
In cuprates, the presence of the strong spin fluctuations 
has often been pointed out especially for underdoped samples. 
For example, we have discussed a possible relevance of 
the strong spin fluctuation to the suppression of $\Tc$ 
in an underdoped region~\cite{Shi00}. 
Therefore, the giant isotope shift in the present result for case (a) 
may explain the experimental results. 
A more detailed discussion on this subject would be presented 
in a separate paper.

In conclusion, 
we have proposed a model of superconductors with 
coexisting interactions of phonon and nonphonon origins. 
In the presence of the Coulomb interaction, 
the extended formulas for $\Tc$ and $\alpha$, 
and a phase diagram are obtained. 
It is found that the inverse isotope effect occurs due to the cancellation 
of the attractive and repulsive interactions, 
where one of them needs to be the phonon interaction, 
while the other the nonphonon interaction 
with a larger effective cutoff energy than the Debye frequency. 
When the phonon interaction is attractive, 
any additional nonphonon pairing interaction reduces 
the value of $\alpha$, 
but does not make $\alpha$ negative. 
It is also found that a strong critical spin fluctuation 
with $\omega_{\rm sf} < \omegaD$ could give rise to 
a giant isotope shift.

The author would like to thank T.~Oguchi for useful discussions.

%%%%%%%%%%%%%%%%%%%%%%%%%%%%%%%%%%%%%%%%%%%%%%%%%%%%%%%%%%%%%%%%%%%%%%%
%%  References                                                       %%
%%%%%%%%%%%%%%%%%%%%%%%%%%%%%%%%%%%%%%%%%%%%%%%%%%%%%%%%%%%%%%%%%%%%%%%

%%%%%%%%%%%%%%%%%%%%%%%%%%%%%%%%%%%%%%%%%%%%%%%%%%%%%%%%%%%%%%%%%%%%%%%


\begin{thebibliography}{99}
%% --------------------------------------------------------------------
%% -- [spin fluctuation mech. for He3. The earliest works.]------------
\bibitem{And73} 
  P.~W.~Anderson and W.~F.~Brinkman: 
  Phys. Rev. Lett. {\bf 30} (1973) 1108;
  S.~Nakajima: Prog. Theor. Phys. {\bf 50} (1973) 1101. 
%% --------------------------------------------------------------------
%% -- [spin fluc. mech. for the heavy fermion. The earliest works.]----
\bibitem{Sca86}
  D.~J.~Scalapino, E.~Loh,~Jr. and J.~E.~Hirsch: 
  Phys. Rev. B {\bf 34} (1986) 8190; 
%% \bibitem{Miy86}
  K.~Miyake, S.~Schmitt-Rink and C.~M.~Varma: 
  Phys. Rev. B {\bf 34} (1986) 6554. 
%% --------------------------------------------------------------------
%% -- [spin fluctuation mech. for the organics. The earliest works.]---
\bibitem{Eme86}
  V.~J.~Emery: Synth. Met. {\bf 13} (1986) 21.
\bibitem{Bea86}
  M.~T.~Beal-Monod, C.~Bourbonnais and V.~J.~Emery: 
  Phys. Rev. B {\bf 34} (1986) 7716. 
\bibitem{Shi89}
  H.~Shimahara: J. Phys. Soc. Jpn. {\bf 58} (1989) 1735; 
  H.~Shimahara: {\it Proceeding of the Physics and Chemistry of
  Organic Superconductors}, eds. G.~Saito and S.~Kagoshima
  (Springer-Verlag, Berlin, Heidelberg, New York, 1990) p.~73. 
%% --------------------------------------------------------------------
%% -- [spin fluctuation mech. for the high Tc. The earliest works.]----
\bibitem{Miy88}
  As works on the spin fluctuation mechanism proposed 
  at an early stage after the discovery of the \hightc cuprates, 
  for example, see 
  [K.~Miyake, T.~Matsuura, K.~Sano and Y.~Nagaoka: 
  J. Phys. Soc. Jpn. {\bf 57} (1988) 722] and ref.~\citen{Shi88}.
%% --------------------------------------------------------------------
%% -- [spin fluc. mech. for the Sr2RuO4. The earliest works.]----------
\bibitem{Ric95}
  T.~M.~Rice and M.~Sigrist: J. Phys. Condens. Matter {\bf 7} (1995) L643; 
%% \bibitem{Maz97}
  I.~I.~Mazin and D.~J.~Singh: 
  Phys. Rev. Lett. {\bf 79} (1997) 733; 
  Phys. Rev. Lett. {\bf 82} (1999) 4324. 
\bibitem{Kle92}
  B.~M.~Klein and R.~E.~Cohen: Phys. Rev. B {\bf 45} (1992) 12405, 
  and references therein. 
\bibitem{Sch96}
  J.~A.~Schlueter {\it et al.}: 
%%  J.~M.~Williams, A.~M.~Kini, U.~Geiser, 
%%  J.~D.~Dudek, M.~E.~Kelly, J.~P.~Flynn, D.~Naumann 
%%  and T.~Roy: 
  Physica C {\bf 265} (1996) 163. 
  %% inverse isotope effect in kL-(ET)2Ag(CF3)4 
\bibitem{Kin96}
  A.~M.~Kini {\it et al.}: 
%%  K.~D.~Carlson, H.~H.~Wang, J.~A.~Schlueter, 
%%  J.~D.~Dudek, S.~A.~Sirchio, U.~Geiser, K.~R.~Lykke 
%%  and J.~M.~Williams: 
  Physica C {\bf 264} (1996) 81, 
  %% inverse isotope effect in k-(ET)2Cu(NCS)2 
\bibitem{IEorg}
  See references in refs.~\citen{Sch96} and \citen{Kin96}. 
\bibitem{Hei86}
  C.~-P.~Heidmann {\it et al.}: 
%%  K.~Andres and D.~Schweitzer: 
  Physica B {\bf 143} (1986) 357; 
  K.~Andres {\it et al.}: 
%%  H.~Schwenk and H.~Veith: 
  Physica B {\bf 143} (1986) 334. 
\bibitem{Osh88}
  K.~Oshima {\it et al.}: 
%%   H.~Urayama, H.~Yamochi and G.~Saito: 
  J. Phys. Soc. Jpn. {\bf 57} (1988) 730. 
\bibitem{Ito91}
  H.~Ito {\it et al.}: 
%%  M.~Watanabe, Y.~Nogami, T.~Ishiguro, T.~Komatsu, G.~Saito 
%%  and N.~Hosoito: 
  J. Phys. Soc. Jpn. {\bf 60} (1991) 3230. 
%% 
\bibitem{Tok91}
  M.~Tokumoto {\it et al.}: 
%%   N.~Kinoshita, Y.~Tanaka and H.~Anzai: 
  J. Phys. Soc. Jpn. {\bf 60} (1991) 1426. 
\bibitem{Mao01}
  Z.~Q.~Mao {\it et al.}: 
%%  Y.~Maeno, Y.~Mori, S.~Sakita, S.~Nimori 
%%  and M.~Udagawa: 
  Phys. Rev. B {\bf 63} (2001) 144514. 
\bibitem{Fou77}
  I.~F.~Foulkes and B.~L.~Gyorffy: Phys. Rev. B {\bf 15} (1977) 1395. 
\bibitem{Shi02a}
  H.~Shimahara and M.~Kohmoto: Europhys. Lett. {\bf 57} (2002) 247. 
\bibitem{Shi02b}
  H.~Shimahara and M.~Kohmoto: Phys. Rev. B {\bf 65} (2002) 174502. 
\bibitem{Shi88}
  H.~Shimahara and S.~Takada: J. Phys. Soc. Jpn. {\bf 57} (1988) 1044. 
\bibitem{Shi03}
  H.~Shimahara: cond-mat/0304516; to be published in J. Phys. Soc. Jpn. 
  {\bf 72}, No.~8 (2003). 
    % Internal transition of superconducting state induced by impurity
    % doping in multiband superconductors
\bibitem{Yam87}
  A similar formula of $\Tc$ has been obtained by Yamaji 
  in a model of $s$-wave pairing superconductivity induced by 
  the pairing interaction mediated by intramolecular vibration modes. 
  [K.~Yamaji: Solid State Commun. {\bf 61} (1987) 413; 
  T.~Ishiguro and K.~Yamaji: {\it Organic Superconductors} 
  (Springer, Berlin, Heidelberg, 1990)]. 
\bibitem{Note03}
  Since $\omegaD \sim 410~{\rm K}$~\cite{Mae97} 
  and $\epsilon_{\rm F} \sim 1.5~{\rm eV}$~\cite{Mae97,Ogu95}, 
  case (a) is realized when 
  $\omega_{\rm sf}/\epsilon_{\rm F} \lsim 0.024$. 
  It corresponds to the peak width of the $\chi(\vq,\omega)$ 
  $\Delta q \lsim 0.024 \times \pi/a$, 
  where $a$ denotes a length scale of the order of the lattice constants. 
  Then, the magnetic correlation length $\xi_{\rm mag}$ needs to satisfy 
  $\xi_{\rm mag} \sim 1/\Delta q \gsim 14 \times a$, 
  which would not be realized in \SrRuO. 
\bibitem{Note02}
  If $\lambda_{\rm ph} < 0$ occurs in \SrRuO, we obtain the reverse conclusion. 
  As far as the author's knowledge, however, there is no study 
  which shows $\lambda_{\rm ph} < 0$ in \SrRuO \hsp{0.25} at the present. 
\bibitem{Kik02}
  N.~Kikugawa and Y.~Maeno: 
  Phys. Rev. Lett. {\bf 89} (2002) 117001. 
\bibitem{Note01}
  Many theories have been proposed for \SrRuO. 
  In addition to ref.~\citen{Ric95}, 
  see also, for example, 
  T.~Kuwabara and M.~Ogata: Phys. Rev. Lett. {\bf 85} (2000) 4586; 
  M.~Sato and M.~Kohmoto: J. Phys. Soc. Jpn. {\bf 69} (2000) 3505; 
  T.~Nomura and K.~Yamada: J. Phys. Soc. Jpn. {\bf 69} (2000) 3678; 
  M.~E.~Zhitomirsky and T.~M.~Rice: Phys. Rev. Lett. {\bf 87} (2001) 057001, 
  and references therein. 
\bibitem{Cra90}
  M.~K.~Crawford {\it et al.}: 
%%  M.~N.~Kunchur, W.~E.~Farneth, E.~M.~McCarron III 
%%  and S.~J.~Poon: 
  Phys. Rev. B {\bf 41} (1990) 282. 
\bibitem{Shi00}
  H.~Shimahara, Y.~Hasegawa and M.~Kohmoto: 
  J. Phys. Soc. Jpn. {\bf 69} (2000) 1598. 
\bibitem{Mae97}
  Y.~Maeno {\it et al.}: 
%%  K.~Yoshida, H.~Hashimoto, S.~Nishizaki, S.~Ikeda, 
%%  M.~Nohara, T.~Fujita, A.~P.~Mackenzie, N.~E.~Hussey, 
%%  J.~G.~Bednortz and F.~Lichtenberg: 
  J. Phys. Soc. Jpn. {\bf 66} (1997) 1405. 
\bibitem{Ogu95}
  T.~Oguchi: Phys. Rev. B {\bf 51} (1995) R1385; 
  D.~J.~Singh: Phys. Rev. B {\bf 52} (1995) R1358. 
\end{thebibliography}
\end{document}